# Measuring the unmeasurable - a project of domestic violence risk prediction and management

Ya-Yun Chen
National Yang-Ming University, Institute of brain science
Taipei, Taiwan
chenyayun9683@gmail.com

Yu-Hsiu Wang
Institute for Information Industry
Taipei, Taiwan
yw2602@columbia.edu

Yi-Shan Hsieh
Yuanta Securities
Taipei, Taiwan
ademon13@163.com

Jing-Tai Ke
Megais Co., Ltd
Taipei, Taiwan
fergus.jingtai@gmail.com

Chia-Kai Liu
DSP, Inc
Taipei, Taiwan
ck@dsp.im

Sue-Chuan Chen
Taipei City Center for Prevention of Domestic Violence and Sexual Assault
Taipei, Taiwan
ha_650080@mail.taipei.gov.tw

T. C. Hsieh*
DSP, Inc
Taipei, Taiwan
johnson@dsp.im

## ABSTRACT

The prevention of domestic violence (DV) have aroused serious concerns in Taiwan because of the disparity between the increasing amount of reported DV cases that doubled over the past decade and the scarcity of social workers. Additionally, a large amount of data was collected when social workers use the predominant case management approach to document case reports information. However, these data were not properly stored or organized.

To improve the efficiency of DV prevention and risk management, we worked with Taipei City Government and utilized the 2015 data from its DV database to perform a spatial pattern analysis of the reports of DV cases to build a DV risk map. However, during our map building process, the issue of confounding bias arose because we were not able to verify if reported cases truly reflected real violence occurrence or were simply false reports from potential victim's neighbors. Therefore, we used the random forest method to build a repeat victimization risk prediction model. The accuracy and F1-measure of our model were 96.3% and 62.8%. This model helped social workers differentiate the risk level of new cases, which further reduced their major workload significantly. To our knowledge, this is the first project that utilized machine learning in DV prevention. The research approach and results of this project not only can improve DV prevention process, but also be applied to other social work or criminal prevention areas.

## 1.INTRODUCTION

Domestic violence (DV) has become a fast-growing problem in Taiwan because of the disproportionate increase in the amount of reported violence cases and the supply of social workers in the past decade. According to the Taiwan Ministry of Health and Welfare, the number of DV cases is 136 thousand in 2015, doubling the amount of the 66 thousand cases in 2005 [1]. Among these cases, intimate partner violence (IPV) cases occupy the largest proportion (53%) in Taiwan.

The importance of DV issues lies in that the violent cases might not only have traumatic impacts on the perpetrator and the victims, they can also affect the mental well-being of their family members [2]. In addition, early experiences of violence in childhood might render the victims into higher chance of repeat victimization or increase the possibilities of them becoming a perpetrator when they turn adults [3]. Adult perpetrators might also carry their violent behavior in their future relationships [4]. While DV problems increased in Taiwan over the past years, numerous data generated from the violence reports have been collected through the predominant case management approach in social work discipline, but were not properly stored and analyzed. Hence, how to utilize existing data to improve the intervention and prevention of DV has attracted the attention of social workers and data scientists.

To improve the efficiency of DV prevention and reduce the workload of social workers, we organized a data scientist team with volunteer experts from diverse disciplines to build a DV risk map and a repeat victimization prediction model in cooperation with Taipei City Center for Prevention of Domestic Violence and Sexual Assault (TPDVPC) under the Data Science for Social Good Fellowship in Taiwan. The ultimate goal of this project was to offer instruments for frontline personnel to efficiently allocate their resources in regions with different violence case types and risk levels and identify high risk victims of repeat victimization when new cases are being reported.

We first performed a spatial pattern analysis of the reported DV cases and built an interactive information risk map. This map visualized DV case reports frequencies and important demographic features of the victims of each DV case type from 456 villages (the fourth-level administrative unit) in Taipei city. However, during our map building process, the issue of confounding bias arose because we were not able to verify if reported cases truly reflected real violence occurrence or were simply false reports from potential victim's neighbors. On the other hand, studies also showed that about half of the violence incidents come from a relatively small number of victims [4, 5]. Thus, a repeat victimization risk prediction model was built subsequently to solve this confounding effect.

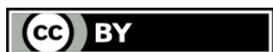

To build the repeat victimization risk prediction model, we focused on building a random forest model with IPV cases using the 2015 data from the database of National Domestic Violence, Sexual Assault, Children and Juvenile Protection Information System at the Ministry of Health and Welfare (MOHW). The result of our prediction model was in line with existing studies-4% of the victims accounted for most repeated IPV case with their reports up to 3 to 7 times in a year. Details of modeling process will be further explained in the next section.

## 2. INTERACTIVE DV PREVENTION RISK MAP

To improve the efficiency of DV prevention and risk management, we proposed an interactive DV risk map. This map visualized case reports frequency of every DV type of the neighborhoods in Taipei City and facilitated the center's implementation of differentiated strategies in response to the specific DV risk levels and types in the diverse communities in Taipei.

### 2.1. DATA SOURCE

In order to build the risk map, we included 8,850 DV cases from the data stored in the National Domestic Violence, Sexual Assault, Children and Juvenile Protection Information System at the MOHW.

### 2.2. DATA PROCESS

- **Classification of event types:** To solve the issues of case types being overly descriptive and complicated, we classified them into four general categories: IPV, children and adolescent protection, elderly protection, and intersibling abuse and others.

- **Location conversion**: We utilized google Maps API to convert address into altitude and latitude and draw the boundaries for districts and villages. Next, each of the district and village in Taipei City administrative unit was labeled with its corresponding numbers of domestic violence reports, violence case types.

- **Map compilation:** Each case type was visualized with its victims' gender ratio, age ratio (0-18, 19-64, 65+), and the numbers of victims with low- and middle- income or with disabilities and mental illness of in terms of percentage. Other reference datasets were collected from the Department of Social Welfare, Taipei City Government and Open Government platform of Data.gov.tw, and was visualized on the interactive map using Python and D3.js.

### 2.3. RESULT

The summary for the data of four DV case types is displayed in Table 1. As shown in the result, IPV cases occupied the largest proportion (52%) among all the case types and the number of female victims was significantly higher than male victims (81% vs. 19%). Compared to the other three case types, children and adolescent protection cases had the largest proportion of victims from low- and middle-income families. In addition, elderly protection and intersibling cases had larger proportions of victims with disabilities or mental illness compared to other two case types.

TABLE I. DATA SUMMARY OF FOUR GENERAL CATEGORIES

| Event type | IPV | children/adolescent | elderly | intersibling and others |
|---|---|---|---|---|
| **Proportion** | 52% | 8% | 6% | 34% |
| **Gender ratio** (Male/Female) | 19/81 | 54/46 | 37/63 | 41/59 |
| **Low- and middle-income** | 8.7% | 24.2% | 9.2% | 12.4% |
| **Disabilities and mental illness** | 2.1% | 1.8% | 3.9% | 4.2% |

A snapshot of our proposed risk information map showed in Fig. 1. The Taipei DV prevention risk map showed that in total 8,850 cases reported in 2015, with higher proportion of female victims (70%). In addition, 75% of the cases fell into the age group between age 18-65.

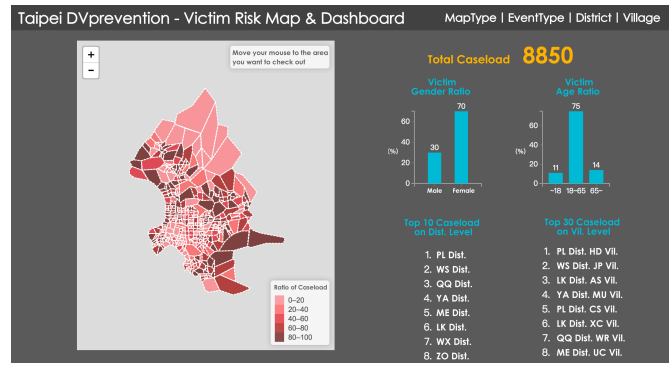

Fig. 1. Snapshot of Main page of DV prevention risk map and dashboard.

The Taipei IPV prevention heatmap (Fig.2) showed that in total 4,617 cases reported in 2015. Females victims occupied even higher proportion (81%) than that of the general DV cases. In addition, a higher proportion of 90% of the cases also fell into the age group between age 18-65.

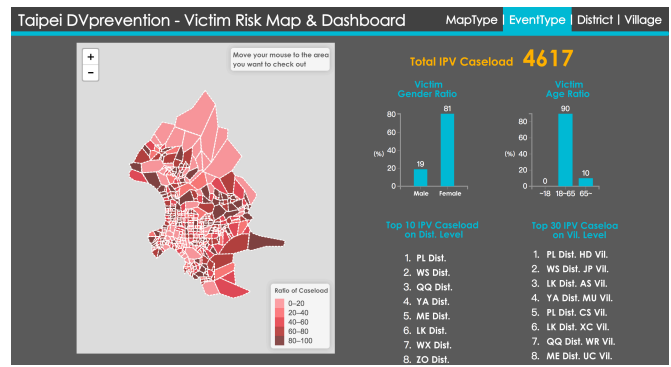

Fig. 2. Snapshot of IPV prevention risk map and dashboard.

In the overview of caseload in each district (Fig. 3), we visualized the numbers of four major case types in bar chart and their corresponding proportion in heat bar. Numbers of victims with the low-and middle- income or disabilities or mental illness are also provided on the district overview page.

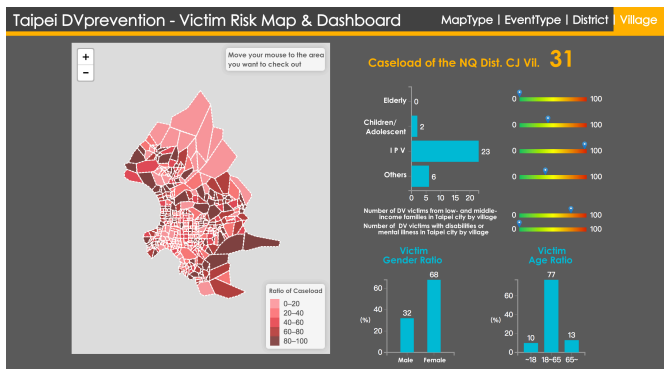

Fig. 3. Snapshot of overview of caseload in each district.

# 3. REPEAT VICTIMIZATION RISK PREDICTION MODEL

The risk map was one of the major results of this project. However, with the display of frequencies into risk map alone could not solve the confounding effect resulting from false positive reports without a cross-validation from repeat report cases. Additionally, past researches also showed that certain features of the violence cases could predict tendency of future repeat victimization [6]. Hence, we built a repeat victimization risk prediction model to produce the prediction data that can solve the confounding issue and strengthen the utility of the original risk map, as well as provide a handy instrument for identifying the repeat victimization risk levels of new cases.

## 3.1. DATA SOURCE

Only the data from the IPV case type are included in our prediction model because they offer the best data quality and occupy up to 52% of the total reports. The data were collected from sources from two forms - the IPV report form with information of the victim and the perpetrator and case-specific details and the Taiwan Intimate Partner Violence Danger Assessment (TIPVDA) form [7]. TIPVDA is a risk assessment instrument assisting front line social workers and other personnel in identifying high risk victims and corresponding intervention strategies. [8]

## 3.2. DATA PROCESS

An exploratory data analysis (EDA) was conducted to summarize the main characteristics of data before building the repeat victimization risk prediction model. The EDA result showed that the distribution of the IPV cases with only one and two report times a year differed from the cases with report times larger than twice (up to 3 to 7) a year.

Next, a new "Response" variable was created, with the value of counts larger than twice set as 1, and the value of counts equal to or smaller than twice set as 0. The records in the "TIPVDA" and "Duration for DV in months" were then grouped into 3 almost equal-sized bins after we sorted the names of variables. Finally, the levels with under 5% frequency were grouped into the same level in the "MAIMED", "OCCUPATION", "EDUCATION", "DISTRICT" variables.

There were 3,759 IPV cases after removing missing value from above variables.

## 3.3. BUILDING MODEL

We hoped to build a classification model to identify the "Response" variable as 1 or 0 (more than or equal to/ smaller than twice IPV case in a year). However, only 4% cases of the Response variable were equal to 1. In other words, it was an imbalanced classification problem. Adoption of traditional classification method might generate a biased result. Therefore, we proposed the following adjusted framework to construct a robust model (Fig. 4).

- **Generate Vacillation Data:** 500 cases are randomly selected into a testing data set.

- **Balanced Resampling:** For the rest of the sample, 500 samples from cases with reports equal to or smaller than twice a year (marked as "O" in Fig.4) and another 500 samples from cases with reports more than twice a year (marked as "X" in Fig. 4) were selected as the training data set.

- **Building Models:** This training data set was then used to build a random forest model with construction of 200 decision trees. The process was repeated 50 times and average as the predictive result.

- **Ensemble**: The process from above steps was repeated 200 times, with its average prediction built into an ensemble random forest model.

- **Result:** The result showed that the average accuracy and recall rate of our proposed model were 96.3% and 88.9%, which indicated a high performance accuracy and sensitivity of our proposed model. Although the model showed a precision and F1-measure of 48.7% and 62.8%, this did not influence the evaluation of the model because the low percentages of precision and F1-measure were consequences of high false positives. In other words, cases with report times less than twice a year were predicted as high risk cases with more than twice reports a year. In social work practices, social workers and their managers tend to take on a more conservative case management approach and thus consider it acceptable for the model to predict a non-high risk case to be a high risk case.

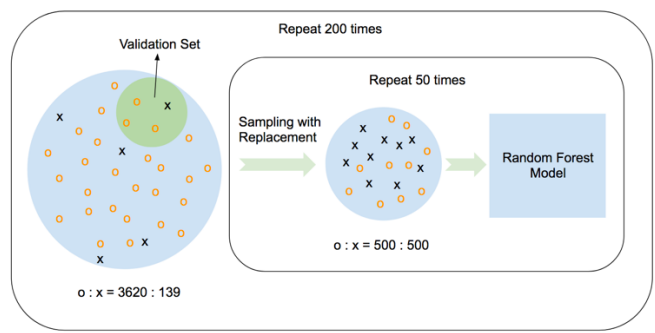

Fig. 4. Methodological framework for building repeat victimization risk model.

# 4. DISCUSSION

This project applied data science solutions to improve the efficiency in utilizing DV data and build new DV prevention strategies. We utilized the 2015 data from Taipei City Government's DV database to build a DV risk map and a IPV repeat victimization prediction model. The following sections outline important insights generated from the project including the potential usages of the risk map and prediction model, solutions to the confounding effect in analysis and suggestions for solving the inter-rater reliability of TIPVDA.

## 4.1. CONTRIBUTION OF THE RISK MAP AND THE REPEAT VICTIMIZATION PREDICATION MODEL

First of all, the risk map visualized complex numeric information in the most straightforward and accessible manner, which facilitated the understanding of the distribution of different case types and various risk levels for its intended users, such as governmental officials, chiefs of village and social workers. In addition, social workers and their managers can identify villages that demand immediate actions by comparing the risk map with areas with concentrated disadvantages, such as villages where public housings are located, and develop differentiated strategies targeting each village according to the specific case types and the demographic features of victims in that village.

On the other hand, the result of our repeat victimization prediction model not only visualized the cases with higher possibilities of repeat victimization in each village in the interactive risk map, it also helped social workers identify the risk level of each case during its first report. Every time a frontline staff input information of a new case into the prediction model, it would analyze the demographic features of the victim and the perpetrator, the forms of the violence being conducted as well as other case details, and instantly gave an output of the risk level indicating the possibility of the victim experiencing repeat victimization in the future. This accessible prevention tool not only stopped potential violence from happening, it also reduced social workers' workload associated with repeat victimization cases significantly.

## 4.2. SOLVING THE CONFOUNDING EFFECT IN THE RISK MAP

As mentioned earlier, a confounding bias might result from the difficulties in identifying the false positive reports, meaning that while a reported case may truly reflect real violence occurrence, it might also come from neighbors who report cases solely based on their own suspicion of violent activities without evidence proving the occurrence of violence. Neglecting this confounding effect might devalue the usability of this risk map. However, while it's impractical to rely on a thorough investigation into each case because of the required investment in time and human resources, we were able to minimize and control the confounding effect by refocusing on the repeat victimization of cases.

## 4.3. THE CONCERN FOR INTER-RATER RELIABILITY OF TIPVDA

The inter-rater reliability of TIPVDA might be reduced because of the inconsistent scores given by frontline personnel with different occupations. One prominent example from our analysis was the scores of TIPVDA given by policemen were lower than the scores of cases evaluated by social workers and staffs in hospitals. The corresponding mean and standard deviation were the following: social worker = 2.71 (0.19), hospital staff = 3.54 (0.19), police = 2.30 (0.18), and the p-value of Mann-Whitney test was smaller than 0.001. This inconsistency in evaluation may result from the insufficient training received by frontline personnel or their unconscious bias toward DV case victims or reporters. Therefore, we suggest future researches on the risk score of DV cases should pay extra attention to the professional background of frontline personnel.

## 4.4. LIMITATIONS AND SUGGESTION

(1) In our exploratory data analysis, positive correlations between high risk cases and items in TIPVDA were found. This was reasonable because TIPVDA is developed for identifying high risk victims [9]. However, we also found a number of slight negative correlations between the variables in repeat victimization cases and items in TIPVDA, suggesting that TIPVDA is not suitable for predicting repeat victimization. In addition to the machine learning-based prediction model built in this project, we suggest that new instrument assessing the risk of repeat victimization cases should also be devised.

(2) While the effect of delaying filling out TIPVDA may not be examined directly, studies should be aware of the fact that the lapse of time between violence occurrence and filling out TIPVDA may lead to a lower risk score since victims' negative perception toward violent experience may be reduced over time [10, 11]. Therefore, social workers and the future investigators should control the factor of when the TIPVDA is filled to ensure higher data collection quality.

## 5. CONCLUSION AND FUTURE WORKS

To the best of our knowledge, this is the first project that utilized machine learning in the DV prevention. We successfully built the DV risk map and used the random forest model to construct the repeat IPV risk prediction model, and has aroused governmental and public's attention to the value of applying data science on DV prevention. Social workers in the TPDVPC have returned positive feedback on using the prediction model and the interactive risk map to facilitate their prevention work. In the future, we will continue to optimize this model by incorporating DV database of other cities, or train a new model for a specific region or city. The result of this project not only help with DV prevention, its approach can also be applied to other criminal prevention areas in the future.

## 6. ACKNOWLEDGEMENTS

We deeply appreciate the partnership of the Department of Social Welfare of Taipei City and Taipei City Center for Prevention of Domestic Violence and Sexual Assault for providing the data and the DSP, Inc. for offering strategic advice and technical support. We would also like to thank the social workers team from TPDVPC, Ying-Yi Chang, Yungji-Chih Huang, Chien-Sheh Chou, Hui-Chuan Lin, Meng-Chuan Hsieh for for contributing DV domain knowledge and assisting us in data correction. We also want to recognize the tireless work of Brian Pan and Tonyq Wang.